\documentclass[aps,pra,twocolumn,superscriptaddress,longbibliography]{revtex4-1}
\usepackage{graphicx}
\usepackage{latexsym}
\usepackage{amssymb}
\usepackage{amsmath}
\usepackage{amsfonts}
\usepackage{upgreek}
\usepackage{float}
\usepackage{bm}
\usepackage{multirow}
\usepackage{color}
\usepackage[T1]{fontenc}
\usepackage{hyperref}
\usepackage{subfigure}
\usepackage{xcolor}

\hypersetup{
colorlinks = true,
linkcolor = [rgb]{0.70,0.13,0.13},
citecolor = [rgb]{0.13,0.55,0.13},
urlcolor  = [rgb]{0.25, 0.41, 0.88}}
\newcommand{\ket}[1]{|#1\rangle}

\newcommand{\braket}[2]{\langle#1|      #2\rangle}

\begin{document}

\title{Quantum criticality in interacting bosonic Kitaev-Hubbard models}

\author{Ya-Nan Wang}
\affiliation{College of Physics, Nanjing University of Aeronautics and Astronautics, Nanjing, 211106, China}
\affiliation{Key Laboratory of Aerospace Information Materials and Physics (Nanjing University of Aeronautics and Astronautics), MIIT, Nanjing 211106, China}

\author{Wen-Long You}
\affiliation{College of Physics, Nanjing University of Aeronautics and Astronautics, Nanjing, 211106, China}
\affiliation{Key Laboratory of Aerospace Information Materials and Physics (Nanjing University of Aeronautics and Astronautics), MIIT, Nanjing 211106, China}

\author{Gaoyong Sun}
\thanks{Corresponding author: gysun@nuaa.edu.cn}
\affiliation{College of Physics, Nanjing University of Aeronautics and Astronautics, Nanjing, 211106, China}
\affiliation{Key Laboratory of Aerospace Information Materials and Physics (Nanjing University of Aeronautics and Astronautics), MIIT, Nanjing 211106, China}

\begin{abstract}
Motivated by recent work on the non-Hermitian skin effect in the bosonic Kitaev-Majorana model, we study the quantum criticality of interacting bosonic Kitaev-Hubbard models
on a chain and a two-leg ladder.
In the hard-core limit, we show exactly that the non-Hermitian skin effect disappears via a transformation from hard-core bosonic models to spin-1/2 models.
We also show that hard-core bosons can engineer the Kitaev interaction, the Dzyaloshinskii-Moriya interaction 
and the compass interaction in the presence of the complex hopping and pairing terms. 
Importantly, quantum criticalities of the chain with a three-body constraint and unconstrained soft-core bosons
are investigated by the density matrix renormalization group method.
This work reveals the effect of many-body interactions on the non-Hermitian skin effect and highlights the
power of bosons with pairing terms as a probe for the engineering of interesting models and quantum phase transitions.

\end{abstract}

\maketitle

\section{Introduction}
The engineering of exotic quantum phases is one of major goals in modern physics.
The fermionic Kitaev chain is a well-known 
quadratic model with $p$-wave superconducting pairing \cite{kitaev2001unpaired},
which exhibits Majorana zero modes localized at the ends of the chain \cite{sarma2015majorana}.
In contrast, for 
the bosonic counterpart, Majorana bosons are forbidden
by the no-go theorems \cite{flynn2021topology}.
In the context of free bosons, another important quantum phase, the Bose-Einstein condensation, appears near absolute zero 
 \cite{anderson1995observation,davis1995bose}.
Recently, the non-Hermitian skin effect \cite{yao2018edge} was proposed and investigated in a one-dimensional bosonic quadratic Hamiltonian with pairing terms \cite{flynn2021topology,mcdonald2018phase, qi2019bosonic, mcdonald2020exponentially, flynn2020deconstructing, flynn2020restoring, yokomizo2021non,chaudhary2021simple}
analogous to the fermionic Kitaev chain.
This phenomenon is particularly interesting as the original bosonic quadratic Hamiltonian is Hermitian, while
the physics of free bosons is characterized by a non-Hermitian Bogoliubov-de Gennes (BdG)
Hamiltonian \cite{mcdonald2018phase,yokomizo2021non}
that leads to the non-Hermitian skin effect.

The non-Hermitian skin effect that corresponds to the localization of bulk states at the boundaries 
reveals the breakdown of the bulk-boundary correspondence and leads to the non-Bloch band theory \cite{yao2018edge}.
The non-Hermitian skin effect has attracted much attention as a unique phenomenon of non-Hermitian systems
without a counterpart in conventional Hermitian models \cite{yao2018edge,lee2016anomalous,kunst2018biorthogonal,xiong2018does,alvarez2018non,gong2018topological,yokomizo2019non,yang2020non, okuma2020topological,zhang2020correspondence,wang2020defective,jiang2020topological,weidemann2020topological,xiao2020non,borgnia2020non,li2020topological,zhang2021observation,zhang2021acoustic,guo2021exact, zhang2022universal,xue2022non}.
The understanding of the interplay between the non-Hermiticity and many-body interactions
is becoming an important research area for many-body systems.
A key issue is whether the non-Hermitian skin effect remains under the many-body interactions.
The study on the non-Hermitian skin effect has previously been considered in 
a few non-Hermitian many-body systems \cite{jin2013scaling,ashida2017parity,herviou2019entanglement,hanai2019non, yamamoto2019theory,hamazaki2019non,chang2020entanglement,matsumoto2020continuous,lee2020many,pan2020non,pan2020interaction,xu2020topological,zhang2020skin,shackleton2020protection, liu2020non,hanai2020critical,mu2020emergent,lee2021many,xi2021classification,yang2021exceptional,sun2022biorthogonal,tang2022dynamical,yang2022hidden,zhangXiangdong2021observation,liang2022observation,suthar2022non, znidaric2022solvable,kawabata2022entanglement}.
To this end, we investigate the non-Hermitian skin effect in the presence of strong interactions in a Hermitian bosonic model with pairing terms \cite{mcdonald2018phase}.

The fermionic Kitaev chain 
can be exactly solved featuring a quantum phase transition in the Ising universality class \cite{kitaev2001unpaired}.
Moreover, the Kitaev chain with many-body interactions has been investigated via various methods as well \cite{fidkowski2011topological,miao2017exact,wouters2018exact,zvyagin2022charging}.
On the other hand, 
low-dimensional bosonic systems have attracted great interest during last decades \cite{greiner2002quantum,lahaye2009physics}.
The competition between hopping terms, interactions and the chemical potential 
induces a large variety of interesting quantum states of matter \cite{bloch2008many}.
However, interacting Bose-Hubbard models with nearest-neighbor pairing terms are less explored \cite{vishveshwara2021z}.

\begin{figure}[t]
\includegraphics[width=8.6cm]{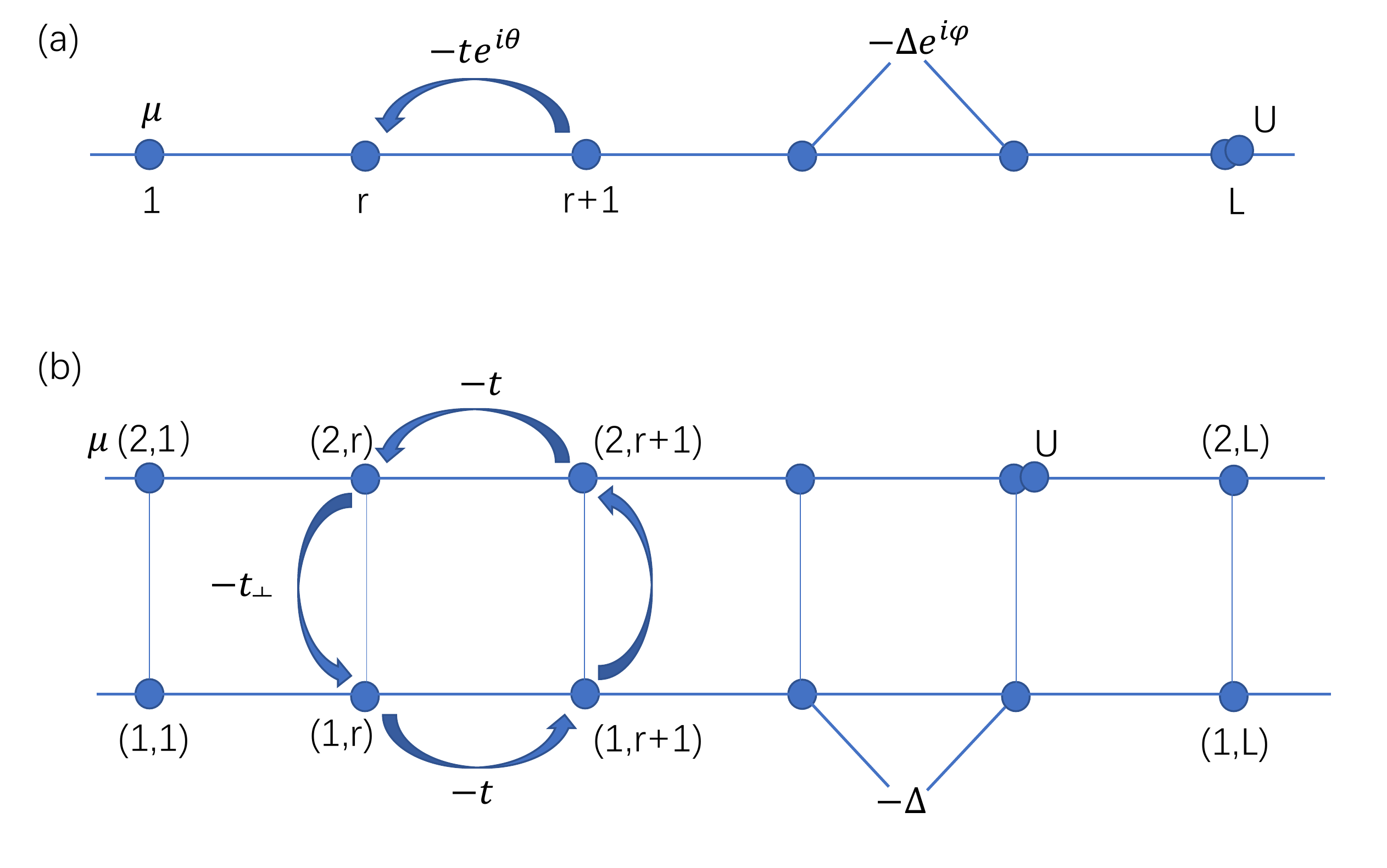} \centering
\caption{Schematic geometry of the BKH models studied in the main text.
(a) One-dimensional chain. (b) Two-leg ladder.
The definitions of parameters $t$, $t_{\perp}$, $\Delta$, $U$, $\mu$, $\theta$ and $\varphi$ for models are shown in the main text.}
\label{model}
\end{figure}

In this paper, we introduce bosonic Kitaev-Hubbard (BKH) models 
on a one-dimensional chain and a two-leg ladder.
The absence of the non-Hermitian skin effect is exactly proved for the chain by mapping hard-core bosons to fermions with the Jordan-Wigner transformation.
Surprisingly, a quantum compass model is found 
at a multi-critical point in which multiple Ising transitions coalesce. 
Most importantly, quantum criticalities and the associated universality are studied by the density matrix renormalization group (DMRG) for the hard-core bosons, three-body constrained bosons, 
and soft-core unconstrained bosons, respectively.
Interesting continuous phase transitions between two ordered phases are discovered for bosons. 

The paper is organized as follows.
In Sec.\ref{sec:chain}, we introduce the BKH model on 
a chain.
In Sec.\ref{sec:skin}, we discuss the non-Hermitian skin effect of the BKH chain for free bosons.
In Sec.\ref{sec:hardcore}, we study the BKH chain in the hard-core limit.
In Sec.\ref{sec:threebody}, we investigate the BKH chain with three-body constraint.
In Sec.\ref{sec:softcore}, we consider the BKH chain for unconstrained soft-core bosons.
In Sec.\ref{sec:ladder}, we present the results of the hard-core BKH model  on a two-leg ladder.
In Sec.\ref{sec:Con}, we summarize our main results.

\section {Bosonic Kitaev-Hubbard Chain}
\label{sec:chain}
The fermionic Kitaev chain is a one-dimensional 
tight-binding model 
with the nearest-neighbor hopping 
and the superconducting $p$-wave pairing terms on each bond \cite{kitaev2001unpaired}.
The corresponding BKH Hamiltonian of a chain as shown in Fig.\ref{model}(a) with complex hopping and pairing amplitudes 
can be defined by \cite{mcdonald2018phase,vishveshwara2021z}
\begin{align}
H=& \sum_{r=1}^{L} (-te^{i \theta}b_r^{\dagger}b_{r+1} - \Delta e^{i \varphi}b_r b_{r+1}+ \text{H.c.}) - \sum_{r=1}^{L} \mu n_r \nonumber \\
&+\sum_{r=1}^{L} \dfrac{U}{2}n_r(n_r-1) + U_3 \sum_{r} n_r (n_{r} -1) (n_{r}-2),
\label{Eq:chain}
\end{align}
where $t \geq 0 $ is the hopping amplitude with a gauge 
phase $\theta$,  
$\Delta \geq 0$ is the pairing amplitude with a phase $\varphi$ that can be gauged out, $\mu$ is the chemical potential,
$U$ and $U_3$ are the strengths of the on-site two-body and three-body interactions.
Here $b_r^{\dagger}$ ($b_r$) and $n_r = b_r^{\dagger}b_{r}$ are the creation (annihilation) and the local particle number operators at the $r$th site.
The periodic boundary condition is imposed as $b_{L+1} = b_{1}$, where $L$ is the system size.
For $U=0$ and $U_3=0$, the system is a free bosonic model, in which the ground state is the Bose-Einstein condensation
at $\Delta=0$. For $U \rightarrow \infty$, the system is a hard-core bosonic model that can be mapped to a spinless free fermionic model
in terms of the Jordan-Wigner transformation. In the following, we will discuss this model in detail.

\section{free bosons}
\label{sec:skin}
Let us firstly study the case of free bosons ($U=0$ and $U_3=0$).
In this section, we will only consider $t > \Delta$ as the opposite regime for $t < \Delta$  is dynamically unstable \cite{mcdonald2018phase}.
For 
$\mu=0$, 
Eq.(\ref{Eq:chain}) can be explicitly written as
\begin{align}
H= \sum_{r=1}^{L} (-te^{i \theta}b_r^{\dagger}b_{r+1} - \Delta e^{i \varphi}b_r b_{r+1} + \text{H.c.}).
\label{Eq:chain2}
\end{align}
Using the Fourier transformation,
\begin{align}
b_r = \frac{1}{\sqrt{L}}\sum_ke^{ikr}b_k,
\end{align}
the model in Eq.(\ref{Eq:chain2}) can be transformed into momentum space with Nambu notations,
\begin{align}
H= \frac{1}{2}\sum_k\left(\begin{array}{cc}
   b_k^{\dagger}  &    b_{-k}  \end{array}\right)
H_{\text{BdG}}(k)
\left(\begin{array}{c}
     b_k  \\
     b_{-k}^{\dagger} \\
\end{array}\right),
\end{align}
in which the BdG Hamiltonian $H_{\text{BdG}}(k)$ reads,
\begin{align}
H_{\text{BdG}}(k) =& -2t \cos \theta \cos k \sigma_0 + 2 t \sin \theta \sin k \sigma_z \nonumber \\
& - 2 \Delta \cos \varphi \cos k \sigma_x - 2 \Delta \sin \varphi \cos k \sigma_y.
\end{align}
Here $\sigma_{x}, \sigma_{y}, \sigma_{z}$ are Pauli matrices and $\sigma_0$ is the identity matrix.
We note that energy eigenvalues of the bosonic Kitaev model in Eq.(\ref{Eq:chain2}) are characterized by a 
modified BdG Hamiltonian,
\begin{align}
H_{\text{BdG}}^{\prime}(k) = \sigma_z H_{\text{BdG}}(k),
\end{align}
because of the bosonic commutation relation \cite{mcdonald2018phase}.
The corresponding energy spectra 
of the bosonic Hamiltonian in Eq.(\ref{Eq:chain2}) are 
derived by solving the characteristic equation,
\begin{align}
\det{[H_{\text{BdG}}^{\prime}(k) - E(k)]}=0,
\end{align}
which yields
\begin{align}
E(k) = 2t \sin \theta \sin k \pm 2\sqrt{t^2 \cos^2 \theta - \Delta^2} \cos k.
\label{Eq:energy}
\end{align}
One easily finds that energy eigenvalues 
in Eq.(\ref{Eq:energy}) are independent on the phase $\varphi$ similar to that of the fermionic Kitaev chain as expected.

\begin{figure}[t]
\includegraphics[width=8.2cm]{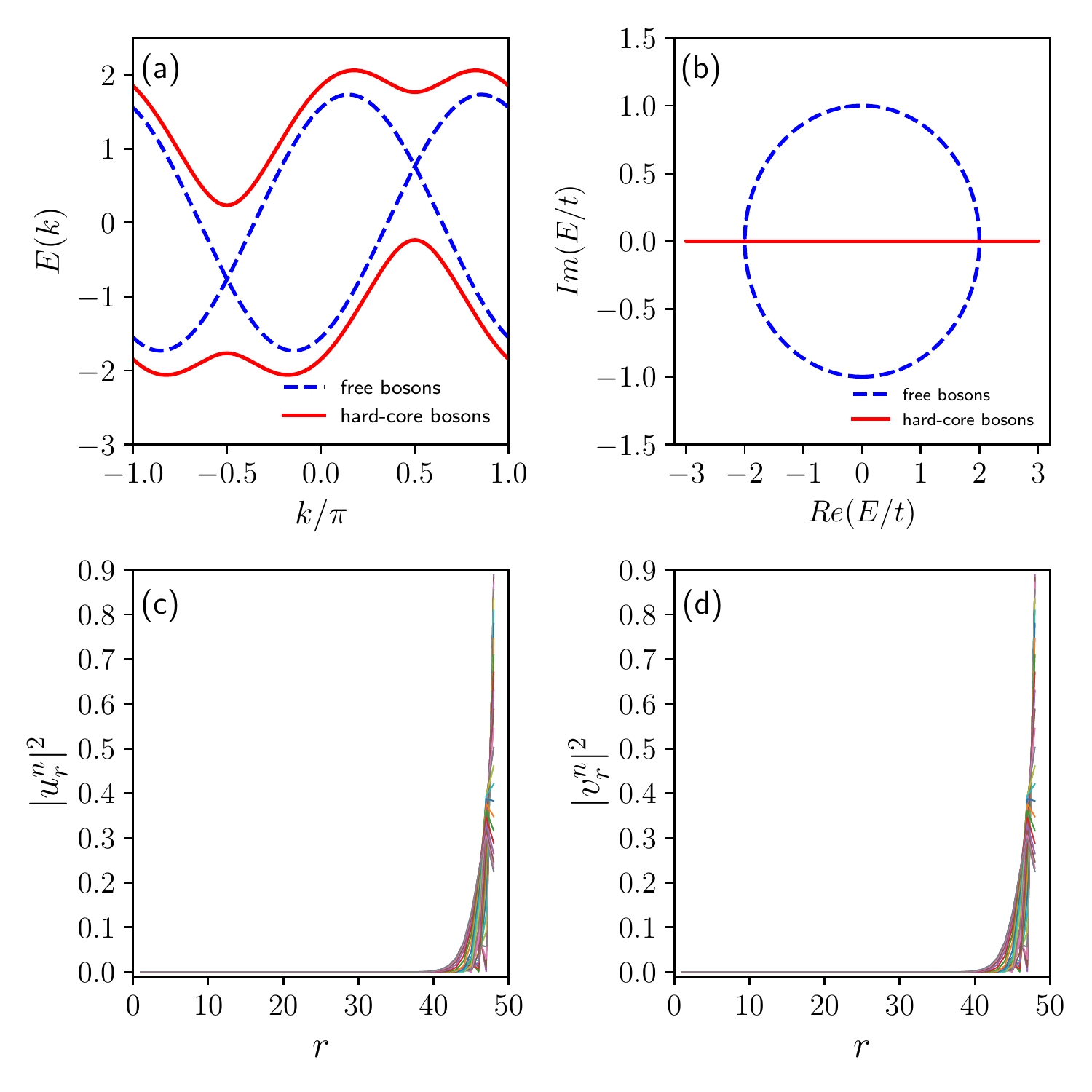} \centering
\caption{
Energy eigenvalues and the eigenstates of the chain.
(a) Energy spectra of the BKH chain for free boson (blue solid line) and hard-core bosons (red dashed line) for $t=1$, $\Delta=1/2$, $\mu=0$ and $\theta=\pi/8$.
(b) Energy spectra of the BKH chain for free bosons (blue ellipse) with $E(k)=2t \sin k \pm 2i \Delta \cos k$
and hard-core bosons (red straight line) with $E(k)=2(t \pm \Delta) \sin k$ 
for $t=1$, $\Delta=1/2$, $\mu=0$ and $\theta=\pi/2$ under periodic boundary conditions.
(c) and (d) Distribution of the coefficients $u_r^{n}$ and $v_r^{n}$ for $L=48$ lattice sites with same parameters in (a) under open boundary conditions.
}
\label{Fig:skin}
\end{figure}

It is worth noting that as
$H_{\text{BdG}}^{\prime}(k)$ can in principle become non-Hermitian,
the non-Hermitian physics may occur despite being a 
Hermitian bosonic model with pairing terms \cite{mcdonald2018phase}.
To perceive it, we firstly consider 
the hopping phase $\theta = \pi/2$. The corresponding energy eigenvalues 
become complex values $E(k)$=$2t \sin k \pm 2i \Delta \cos k$ as shown in Fig.\ref{Fig:skin}(b), indicating the occurrence 
of the non-Hermitian skin effect \cite{mcdonald2018phase},
where all bulk states are localized at the boundaries [cf. Fig.\ref{Fig:skin}(c) and Fig.\ref{Fig:skin}(d)].
For an open chain, the Hamiltonian can also be diagonalized by the Bogoliubov transformation \cite{mcdonald2018phase},
\begin{align}
H = \sum_{n} E_n \beta^{\dagger}_{n} \beta_{n}.
\end{align}
Here, $E_n=2 \sqrt{t^2 - \Delta^2}\cos k_n$, and the $\beta_{n}= \sum_{r} u_{r}^{n}b_{r} - v_{r}^{n}b^{\dagger}_{r}$ is the quasiparticle with the coefficients $u_r^{n}$ and $v_r^{n}$ given by 
\cite{mcdonald2018phase},
\begin{align}
u_{r}^{n} = \sqrt{\frac{2}{N+1}} i^{-r} \sin(k_n r) \cosh(\gamma r), \\
v_{r}^{n} = \sqrt{\frac{2}{N+1}} i^{-r} \sin(k_n r) \sinh(\gamma r),
\end{align}
where $k_n = n \pi/(L+1)$  ($n=1, \cdots, L)$ and $e^{2 \gamma} = (t +\Delta)/(t -\Delta)$.  
The coefficients $u_r^{n}$ and $v_r^{n}$ shown in Fig.\ref{Fig:skin}(c) and Fig.\ref{Fig:skin}(d) 
represent the particle and antiparticle parts of the wave function \cite{mcdonald2018phase},
which are localized at the boundary of the chain as expected.

In contrast, in the case of $t \cos \theta > \Delta$, the system in Eq.(\ref{Eq:chain2}) has real energy spectra 
 $E(k) =  A \cos (k - \psi)$ 
and $E(k) = - A \cos (k + \psi)$ [cf. Fig.\ref{Fig:skin}(a)],
with $A=2 \sqrt{t^2-\Delta^2}$ and $\sin \psi = t \sin \theta / \sqrt{t^2-\Delta^2}$.
Consequently, the choice of hopping phase $\theta$ plays a key role in the emergence of the non-Hermitian skin effect \cite{mcdonald2018phase}. 
By comparison, the hopping phase $\theta$ is less considered in the original fermionic Kitaev chain as it might merely shift phase boundaries of the Ising transition.
We will explore it in detail in the next section.

\begin{figure}[t]
\includegraphics[width=8.6cm]{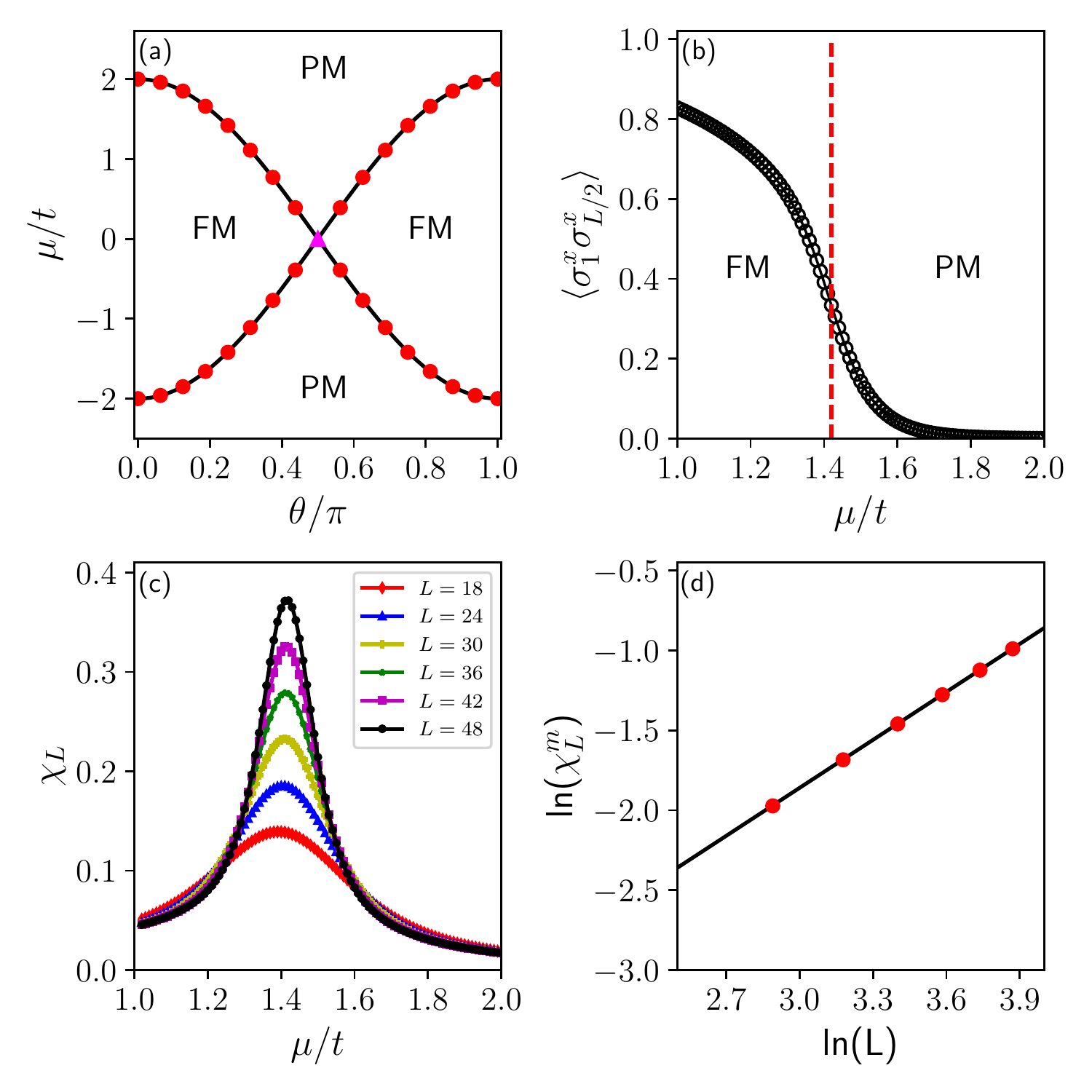} \centering
\caption{Hard-core BKH chain with $t=\Delta=1$ for periodic boundary conditions.
(a) The phase diagram with respect to the hopping phase $\theta / \pi$ and the chemical potential $\mu/t$, where the solid (dotted) lines are obtained by analytical (DMRG) calculations.
The peak triangle symbol at $\theta=\pi/2$, $\mu /t=0$ denotes the quantum compass model.
(b) The correlation function $\langle \sigma_{1}^{x} \sigma_{L/2}^{x} \rangle$ as a function of $\mu/t$ at $\theta=\pi/4$. 
(c) The fidelity susceptibility per site $\chi_{L}$ with respect to $\mu /t$ for $L=18$ to $L=48$ sites (from bottom to top) with $\theta=\pi/4$. 
(d) The correlation-length critical exponent $\nu=0.9969 \pm 0.0047$ is derived from the finite-size scaling of the maximum fidelity susceptibility.
}
\label{chainphase}
\end{figure}

\section{hard-core bosons}
\label{sec:hardcore}
In the strong two-body interaction limit $U \rightarrow \infty$ (hard-core bosons), the Hilbert space of bosons at each local site can be truncated into two states $\ket{0}$ and $\ket{1}$, 
which can be treated as effective spin-1/2 states.
The transformations between bosonic operators $b_{r}^{\dagger}$, $b_{r}$
and spin operators $\sigma_{r}^{+}$,  $\sigma_{r}^{-}$, $\sigma_{r}^{z}$, can be written as
$\sigma_{r}^{+} = b_{r}$, $\sigma_{r}^{-} = b_{r}^{\dagger}$, $\sigma_{r}^{z} = 1-2b_{r}^{\dagger}b_{r}$
by identifying $\ket{0} \rightarrow \ket{\uparrow}$ and $\ket{1} \rightarrow \ket{\downarrow}$.
In this respect, the Hamiltonian in Eq.(\ref{Eq:chain}) can be mapped onto a transverse field spin chain as,
\begin{align}
H_{S}&=&-\frac{1}{2}\sum_{r=1}^{L} \left[ (t \cos \theta + \Delta) \sigma_{r}^{x} \sigma_{r+1}^{x} + (t \cos \theta - \Delta) \sigma_{r}^{y} \sigma_{r+1}^{y} \right] \nonumber \\
&& + \frac{1}{2}\sum_{r=1}^{L} t \sin \theta (\sigma_{r}^{x} \sigma_{r+1}^{y}\!-\!\sigma_{r}^{y} \sigma_{r+1}^{x})\! -\!\frac{1}{2} \sum_{r=1}^{L} \mu (1\!-\!\sigma_{r}^{z}), \quad
\label{Eq:spinchain}
\end{align}
where the raising (lowering) operators $\sigma_{r}^{\pm}=
(\sigma_{r}^{x} \pm i\sigma_{r}^{y})/2$ have been used.
Here, we obtain the anisotropic XY model with the Dzyaloshinskii-Moriya (DM) interaction.
The effective spin model in Eq.(\ref{Eq:spinchain}) can be transformed into the fermionic Kitaev chain,
\begin{align}
H_{F} =   \sum_r(-te^{i\theta}c_{r}^{\dagger}c_{r+1} - \Delta c_{r}c_{r+1} + \text{H.c.}) - \sum_r\mu c_r^{\dagger}c_r, \quad
\end{align}
under the Jordan-Wigner transformation,
\begin{align}
\sigma_r^{+}=& \prod_{i=1}^{r-1}(1-2c_i^{\dagger}c_i)c_r, & 
\sigma_r^-=  \prod_{i=1}^{r-1}(1-2 c_i^{\dagger}c_i)c_r^{\dagger}, \nonumber \\
\sigma_r^z=& 1-2 c_r^{\dagger}c_r.
\end{align}
When $\theta = 0$, the fermionic Kitaev chain exhibits topological Majorana bound states in the regime $-2 < \mu/t < 2$.
The phase transition between the topological phase and the trivial phase 
occurring at $\mu/t = \pm 2$ belongs to the Ising universality class.

As the hopping phase $\theta$ in the fermionic Kitaev chain cannot be gauged out in the presence of the pairing term $\Delta$,
it is expected that the phase $\theta$ would modify the phase diagram.
The energy spectrum of the fermionic Kitaev chain for an arbitrary phase $\theta$ is derived as,
\begin{align}
E(k) = 2t \sin \theta \sin k \pm \sqrt{(\mu + 2t \cos \theta \cos k)^2 + (2 \Delta \sin k)^2}.
\label{Eq:Kitaevchain}
\end{align}
The critical point is obtained as $\mu_c = \pm 2t \cos \theta$ for $\Delta > t \sin \theta$ by solving the equation $E(k)=0$ in Eq.(\ref{Eq:Kitaevchain}) \cite{yi2019criticality,zhao2022characterizing}. 

In order to confirm our analysis, we perform the DMRG \cite{white1992density,schollwock2011density} calculations
for the spin chain in Eq.(\ref{Eq:spinchain}) with $t=\Delta$
under periodic boundary conditions. The critical values $\mu_c$ and the correlation-length critical exponent $\nu$ are obtained
by the ground-state fidelity susceptibility \cite{gu2010fidelity,you2007fidelity,sun2017fidelity},
\begin{align}
\chi_{F} = \lim_{\delta \lambda \rightarrow 0} \frac{-2\ln F(\lambda,\lambda+\delta \lambda)}{(\delta \lambda)^2}.
\end{align}
Here $F(\lambda, \lambda+\delta \lambda)=|\braket{\psi(\lambda)}{\psi(\lambda+\delta \lambda)}|$ is the ground-state fidelity
with a control parameter $\lambda \equiv \mu$.
For second-order phase transitions, the fidelity susceptibility per site $\chi_L \equiv \chi_F / L$ 
near the critical point in one dimension scales as \cite{gu2010fidelity,you2007fidelity,sun2017fidelity},
\begin{align}
\chi_{L} \propto L^{2 / \nu -1}.
\end{align}

\begin{figure}[t]
\includegraphics[width=8.6cm]{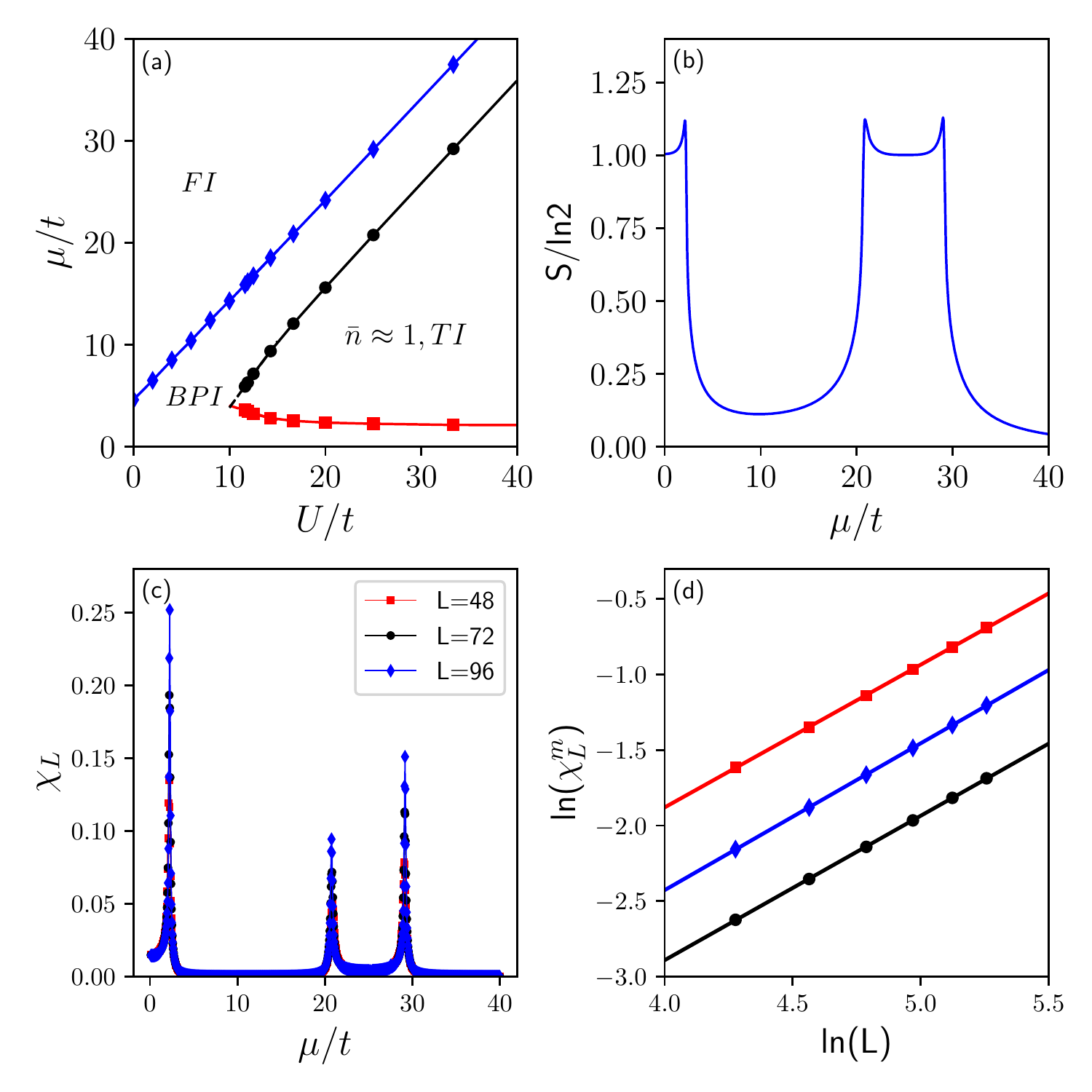} \centering
\caption{ Phase diagram of the BKH chain with three-body constraint at $t=\Delta=1$ under open boundary conditions.
(a) The phase diagram as a function of the interaction $U/t$ and the chemical potential $\mu/t$ for $L=96$ lattice sites.
(b) The half-chain entanglement entropy $S$ with respect to $\mu/t$ for $U=25$ and $n_{\text{c}}=2$.
(c) The fidelity susceptibility per site $\chi_{L}$ with respect to $\mu /t$ for $L=48, 72, 96$ sites. 
(d) Correlation-length critical exponents $\nu=1.0289 \pm 0.0022$ (red square), $\nu=1.0246 \pm 0.0017$ (black circle) and $\nu=1.0180 \pm 0.0011$ (blue diamond)
are derived from the finite-size scaling of the maximum fidelity susceptibility (from left to the right) up to $L=192$ sites.
}
\label{n2phase}
\end{figure}

The phase diagram shown in Fig.\ref{chainphase}(a) is
computed from the fidelity susceptibility [cf. Fig.\ref{chainphase}(c)] for $t=\Delta=1$, where numerical results are consistent with analytical values.
The phase transition remains the Ising transition from the ferromagnetic (FM) phase
to the paramagnetic (PM) phase [cf. Fig.\ref{chainphase}(b)] with the correlation-length critical exponent $\nu=1$ [cf. Fig.\ref{chainphase}(d)] for a finite $\theta$.
We note that the hard-core BKH chain with periodic boundary conditions
is exactly equivalent to the fermionic Kitaev chain with antiperiodic boundary conditions on a finite system.
Interestingly, two Ising transition lines merge at $\mu=0$, $\theta = \pi/2$, implying that this point might be a multi-critical point.
We show that this special point corresponds to the celebrated compass model with $2^{L/2-1}$-fold
degenerate ground states \cite{you2008quantum,eriksson2009multicriticality,you2014quantum,wu2019exact}.
To be more explicit, the Hamiltonian of the hard-core BKH chain with $t=\Delta$, $\theta=\pi/2$ and $\mu=0$ is given by
\begin{align}
H_{S}^{\prime} =& -\frac{\Delta} {2}  \sum_{r=1}^{L}  (\sigma_{r}^{x} + \sigma_{r}^{y})  (\sigma_{r+1}^{x} -\sigma_{r+1}^{y}) \nonumber \\
=& -\Delta \sum_{r=1}^{L} \tilde{\sigma}_{r}^{y} \tilde{\sigma}_{r+1}^{x}.
\label{Eq:spinhalf}
\end{align}
Here, we introduce new spin operators $\tilde{\sigma}_{r}^{x} = (\sigma_{r}^{x} - \sigma_{r}^{y})/\sqrt{2}$ and $\tilde{\sigma}_{r}^{y} = (\sigma_{r}^{x} + \sigma_{r}^{y})/\sqrt{2}$.
Performing a transformation only for the even sites $\tilde{\sigma}_{2r}^{y} \rightarrow \tilde{\sigma}_{2r}^{x}$, $\tilde{\sigma}_{2r}^{x} \rightarrow \tilde{\sigma}_{2r}^{y}$
and $\tilde{\sigma}_{2r}^{z} \rightarrow \tilde{\sigma}_{2r}^{z}$, we arrive at the following compass model \cite{eriksson2009multicriticality,you2014quantum},
\begin{align}
H_{S}^{\prime} = -\Delta \sum_{r=1}^{L} \left( \tilde{\sigma}_{2r-1}^{y} \tilde{\sigma}_{2r}^{y} + \tilde{\sigma}_{2r}^{x} \tilde{\sigma}_{2r+1}^{x} \right)
\label{Eq:compass}
\end{align}
The ground-state degeneracy of the hard-core BKH chain in Eq.(\ref{Eq:chain}) at $t=\Delta$, $\theta=\pi/2$ and $\mu=0$
is numerically verified for small systems using the exact diagonalization, which is consistent with the analytical result \cite{you2008quantum,eriksson2009multicriticality,you2014quantum,wu2019exact}
of the compass model in Eq.(\ref{Eq:compass}). 

As the energy eigenvalues $E(k)$ in Eq.(\ref{Eq:Kitaevchain}) are always real [cf. Fig.\ref{Fig:skin}(a) and Fig.\ref{Fig:skin}(b)], the non-Hermitian skin effect cannot occur for hard-core bosons.
This is because hard-core bosons obey fermionic commutation relation instead of the original bosonic commutation relation, from which the non-Hermitian skin effect may appear. 
We note that non-Hermitian skin effect of the bosonic BdG Hamiltonian may be instability against infinitesimal perturbations  (i.e. the chemical potential $\mu$)  \cite{yokomizo2021non}.
It would be more interesting to study the impact of the finite 
interaction $U$ on the non-Hermitian skin effect. 
However, the huge local Hilbert space of bosons on each lattice site retards 
the possibility of exact or numerical simulations of full spectrum.  
Besides, it is still an open question for 
the characterization of non-Hermitian skin effect in non-integrable many-body systems.
The study on the non-Hermitian skin effect for arbitrary finite $U$ is left for future study. 
In the following, we will mainly focus on the quantum criticalities of the BKH model with interactions.

\section{three-body constrained bosons}
\label{sec:threebody}
The interaction between bosons can be tuned by Feshbach resonances \cite{chin2010feshbach} in experiments.
In fact, the hard-core bosons may be realized approximately by a 
sufficiently strong on-site repulsion \cite{de2019observation} instead of the infinite $U$. 
Consequently, we expect similar phase transitions discussed above for the hard-core bosons persist for the large $U$. 
First, let us consider the three-body constraint due to the Zeno-like effect \cite{daley2009atomic,greschner2013ultracold}, 
where more than double occupancy is suppressed.
The three-body constraint for the model in Eq.(\ref{Eq:chain}) can be achieved in the strong three-body interaction 
limit $U_3 \rightarrow \infty$\cite{greschner2013ultracold}. 

The full phase diagram is presented in Fig.\ref{n2phase}(a) for the Hamiltonian in Eq.(\ref{Eq:chain}) with three-body constraint ($U_3 \rightarrow \infty$), which is obtained
 by performing the DMRG calculations through the fidelity susceptibility shown in Fig.\ref{n2phase}(c)
with the maximal on-site occupancy $n_{\text{max}}=2$ for $t=\Delta=1$ and $\theta=\varphi=0$.
Meanwhile, the entanglement entropy $S=-\text{tr} \rho_{A} \text{log} \rho_{A}$ is also calculated, where $ \rho_{A} = \text{tr}_{B} \rho$ is the reduced density matrix of the subsystem.
In the large interaction limit ($U \gg t$), the ground state of the system
is the gapped bond pairing insulator (BPI) phase at $\mu =0$. We note that the BPI phase corresponds to the FM phase of the spin model 
or the Majorana topological phase of the fermionic Kitaev chain in the hard-core limit ($U \rightarrow \infty$). 
The half-chain entanglement entropy of the BPI phase is $S \approx \ln2$ [cf. Fig.\ref{n2phase}(b)]. 
Increasing the chemical potential $\mu$, the system undergoes a phase transition from the BPI phase to trivial insulator phase (TI) at the filling $\bar{n} \approx 1$. 
Here, the $\bar{n} \approx 1$ TI phase with the half-chain entanglement entropy $S \approx 0$  [cf. Fig.\ref{n2phase}(b)] 
is related to the FM phase (or the trivial phase) in the spin model (or in the fermionic Kitaev chain) 
in the hard-core limit. The BPI phase revives between the $\bar{n} \approx 1$ TI phase and the fully filled insulator (FI) phase 
(vacuum phase $\bar{n} \approx 2$) by a further increase of $\mu$. Interestingly, we find the direct phase transition between the BPI phase and the FI phase 
even without the  on-site two-body interaction, i.e., $U=0$. All phase transitions are found to 
fall into the Ising universality class [cf. Fig.\ref{n2phase}(d)].

\begin{figure}[t]
\includegraphics[width=8.6cm]{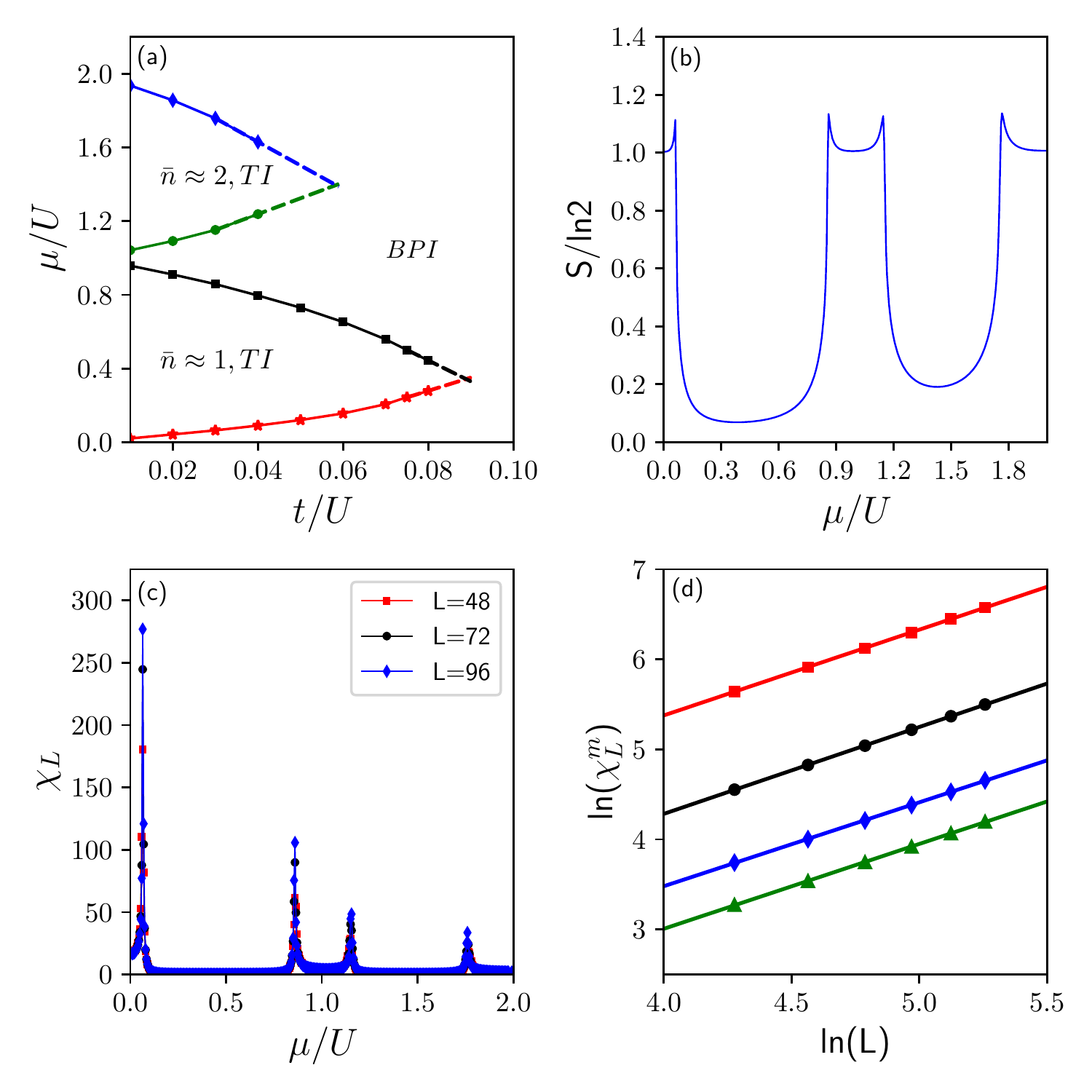} \centering
\caption{ Phase diagram of the BKH chain for soft-core bosons at $U=1$ under open boundary conditions.
(a) The phase diagram as a function of the interaction $t/U$ and the chemical potential $\mu/U$ for $L=96$ lattice sites.
(b) The half-chain entanglement entropy $S$ with respect to $\mu/U$ for $t/U=0.03$ and $n_{\text{c}}=5$.
(c) The fidelity susceptibility per site $\chi_{L}$ with respect to $\mu /t$ for $L=48, 72, 96$ sites. 
(d) Correlation-length critical exponents $\nu=1.0234 \pm 0.0050$ (red square),  $\nu=1.0173 \pm 0.0036$ (black circle), $\nu=1.0338 \pm 0.0072$ (blue diamond), $\nu=1.0283 \pm 0.0060$ (green triangle)
are derived from the finite-size scaling of the maximum fidelity susceptibility (from left to the right) up to $L=192$ sites.
}
\label{softcorephase}
\end{figure}

\section{Soft-core bosons}
\label{sec:softcore}
Next, let us now discuss the soft-core regime, in which the three-body constraint is removed ($U_3=0$). 
When the pairing term is absent ($\Delta=0$), the model reduces to 
the usual Bose-Hubbard model.  On one hand, for certain chemical potential $\mu$, the competition between the kinetic energy $t$ and the 
repulsive energy $U$ leads to the transition between the Mott insulator ($t \ll U$) 
and the superfluid ($t \gg U$) 
\cite{greiner2002quantum}.
When the kinetic term $t$ is replaced by the pairing term $\Delta$, (that is $t=0$ but $\Delta \neq 0$), it is argued that no phase transition happens \cite{correa2013bose}. 
Instead, the system may undergo a crossover from the insulating phase ($\Delta \ll U$) to the squeezed phase ($\Delta \gg U$) \cite{correa2013bose}.

Interestingly, we find a direct phase transition between the BPI phase and the TI phases [cf. Fig.\ref{softcorephase}(a)] in the BKH model.
This is because the U(1) symmetry of the original Bose-Hubbard model is reduced to the $\mathbb{Z}_2$ symmetry 
when both the kinetic term $t \neq 0$ and the pairing term $\Delta \neq 0$.
A phase transition can in principle occur between two gapped phases.
We obtain the global phase diagram presented in Fig.\ref{softcorephase}(a) with TI lobes from the fidelity susceptibility [cf. Fig.\ref{softcorephase}(c)] by the DMRG with truncated on-site occupancy  $n_{\text{max}}=5$,
which is analogous to the original Bose-Hubbard model. 
The nature of the transition between the BPI phase and the TI phase remains the Ising transition [cf. Fig.\ref{softcorephase}(d)], which can also be described by the entanglement entropy [cf. Fig.\ref{softcorephase}(b)] 
as discussed above. In the strong interaction limit $U \gg t,\Delta$, the results are entirely consistent with the analysis in the hard-core 
limit
[cf. Fig.\ref{softcorephase}(a)].

\section{hard-core Bosonic Kitaev-Hubbard ladder}
\label{sec:ladder}
\begin{figure}[t]
\includegraphics[width=8.6cm]{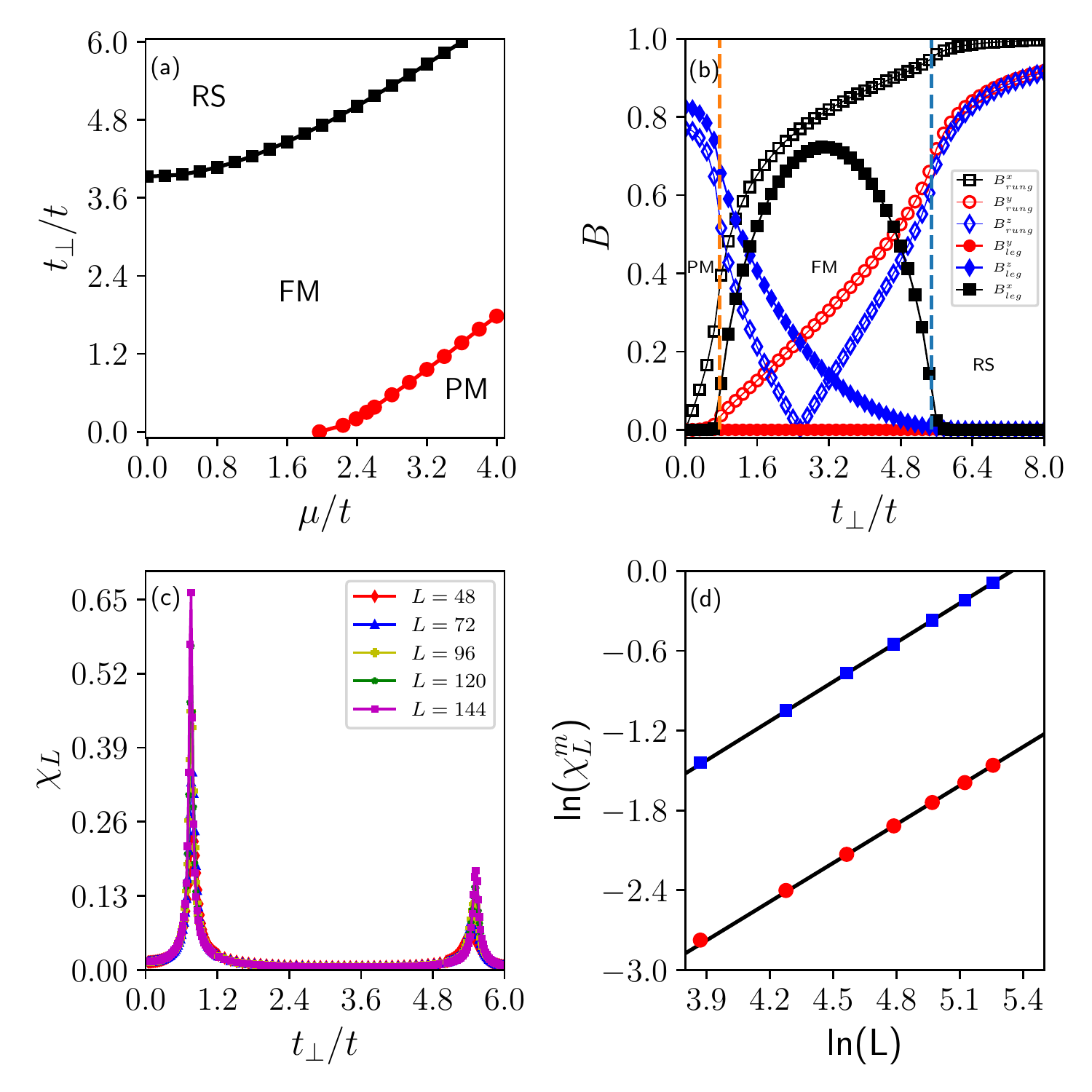} \centering
\caption{ Hard-core BKH ladder with $t=\Delta=1$ under open boundary conditions.
(a) The phase diagram with respect to the rung hopping $t_{\perp}/t$ and the chemical potential $\mu/t$ for $L=96$ rungs.
(b) The bond order parameters $B_{\text{leg}}^{\alpha}=\langle \sigma_{1,L/2}^{\alpha} \sigma_{1,L/2+1}^{\alpha} \rangle$ along the leg 
and $B_{\text{rung}}^{\alpha}=\langle \sigma_{1,L/2}^{\alpha} \sigma_{2,L/2}^{\alpha} \rangle$ along the rung as a function of $t_{\perp} /t$ at $\mu/t=3$ and $\theta=\pi/4$, where $\alpha=x$, $y$, $z$.
(c) The fidelity susceptibility per site $\chi_{L}$ with respect to $t_{\perp}/t$ for $L=48$ to $L=144$ rungs (from bottom to top) with $\mu/t=3$ and $\theta=\pi/4$.
(d) The finite-size scaling of the maximum fidelity susceptibility.
The correlation-length critical exponents $\nu=1.0112 \pm 0.0034$ (blue square symbols) and $\nu= 1.0240 \pm 0.0088$ (red circles) are obtained from fitting first and second peaks
up to $L=192$ rungs.
}
\label{ladderphase}
\end{figure}

Having discussed the interacting bosonic Hubbard chain with pairing terms, we now turn to a two-leg ladder as shown in Fig.\ref{Eq:chain}(b),
where the inter-leg hopping is included. For the sake of simplicity,
we will consider only the hard-core bosons. And the hopping integrals are assumed to be real
as the hopping phase $\theta$ would merely shift phase boundaries as shown in Fig.\ref{chainphase}(a)
except for the special point $\theta = \pi/2$.
The corresponding Hamiltonian of the BKH model with the real hopping matrix $t$ 
and the real bosonic pairing $\Delta$ on a two-leg ladder is then given by, 
\begin{align}
H_L =& -\sum_{l=1,2;r=1}^{L} ( t b_{l,r}^{\dagger}b_{l,r+1}+\Delta b_{l,r}b_{l,r+1} + \text{H.c.} ) \nonumber \\
& -\sum_{r=1}^{L}(t_{\perp} b_{1,r}^{\dagger}b_{2,r} + \text{H.c.}) - \sum_{l=1,2;r=1}^{L}\mu n_{l,r} \nonumber \\ 
& +\sum_{l=1,2;r=1}^{L} \dfrac{U}{2}n_{l,r}(n_{l,r} - 1),
\end{align}
where $b_{l,r}^{\dagger}$ is the bosonic creation operator at the $l$th leg and the $r$th rung,
$t$ and $t_{\perp}$ are hopping matrix elements along legs and rungs.
Here, the bosons are assumed to be paired only on the legs.
In order to understand the phase diagram, we map the hard-core BKH ladder ($U \rightarrow \infty$) onto the spin ladder, 
whose Hamiltonian is described by
\begin{align}
H_{LS} =&  -\frac{1}{2}\sum_{l=1,2;r=1}^{L} \left[ (t+\Delta) \sigma_{l,r}^{x} \sigma_{l,r+1}^{x} + (t-\Delta) \sigma_{l,r}^{y} \sigma_{l,r+1}^{y} \right] \nonumber \\
&- \frac{1}{2}\sum_{r=1}^{L} t_{\perp} (\sigma_{1,r}^{x} \sigma_{2,r}^{x} + \sigma_{1,r}^{y} \sigma_{2,r}^{y}) - \frac{1}{2} \sum_{r=1}^{L} \mu (1-\sigma_{l,r}^{z}).
\end{align}
When $t_{\perp}=0$, the system consists of two decoupled bosonic Kitaev chain as we discussed in Sec.\ref{sec:hardcore}. The ground state is
an XY phase with a U(1) symmetry at $\Delta=0$.
When a nonzero $\Delta > 0$ is included, the U(1) symmetry is reduced to the $\mathbb{Z}_2$ symmetry. 
The ground state becomes a FM phase along the $x$-direction.
In the opposite limit $t_{\perp} \rightarrow \infty$, the ground state of the system 
evolves into a rung-singlet (RS) phase \cite{vekua2003phase,lauchli2003phase,hijii2005phase}.
In between, a direct continuous second-order phase transition can take place between
the FM phase and the RS phase \cite{vekua2003phase,lauchli2003phase,hijii2005phase}.
To verify it, we perform the DMRG calculation with open boundary conditions up to $L=192$ rungs. The full phase diagram is presented in 
Fig.\ref{ladderphase}(a), 
which is obtained from the fidelity susceptibility in Fig.\ref{ladderphase}(c). 
In addition to the phase transition between the FM phase and the PM 
phase,
the system exhibits another second-order phase transition between the FM phase and the RS phase that can be 
captured by bond order parameters  $B=(B_{\text{leg}},B_{\text{rung}})$
exhibited in Fig.\ref{ladderphase}(b). 
The nature of the phase transition as shown in Fig.\ref{ladderphase}(d) belongs to the Ising universality class as well.

Furthermore, when many-body interactions $H_{\text{V}} = V \sum\limits_{r=1}^{L} (n_{1,r} - \frac{1}{2})(n_{2,r} - \frac{1}{2})$ of hard-core bosons along rungs
is considered, the system can be mapped to a spin-1/2 Ising ladder,
\begin{align}
H_{LI} = H_{LS} + \frac{1}{4} V \sum_{r=1}^{L} \sigma_{1,r}^{z} \sigma_{2,r}^{z},
\end{align}
with the interaction along the $z$-direction.
In the case of $t_{\perp}=0$, the model is decoupled into two interacting fermionic Kitaev chains \cite{herviou2016phase}.
Especially, the system reduces to the quantum compass ladder that can be solved exactly at $t_{\perp}=0$, $\Delta=1$ and $\mu=0$ \cite{brzezicki2009exact}.
Hence, hard-core bosons on a ladder with pairing terms also offer a simple way to mimic various interesting many-body spin models.

\section {Conclusion}
\label{sec:Con}
In summary, we have studied the phase diagram 
and associated phase transitions for BKH models on a chain and a two-leg ladder.
We show that one-dimensional BKH chain in the hard-core boson limit is identical to the fermionic Kitaev chain
and hereby declare that the non-Hermitian skin effect should vanish.

Moreover, we reveal that in the presence of the hopping phase the Dzyaloshinskii-Moriya interactions can be engineered for hard-core bosons.
The ground state of the system remains the doubly degenerate Ising phase with the $\mathbb{Z}_2$ symmetry for any $\theta$ $ (\theta \neq \pi/2)$ in the case $\Delta > t \sin \theta$, 
while the non-Hermitian skin effect can emerge for free bosons by tuning the phase $\theta$.
Interestingly, the effective compass model can be realized characteristic of a $2^{L/2-1}$-fold degenerate ground state at $\theta=\pi/2$.
It is surprising that the soft-core BKH chain exhibits a direct phase transition from a TI phase to the BPI phase, which are argued to be a crossover with only the paring term $\Delta \neq 0$.
For the two-leg ladder, we find a continuous phase transition between the ordered FM phase and the RS phase falling into the Ising universality class as ever.
More relevant interesting models are finally discussed.
It would be more interesting to investigate two-dimensional many-body systems to explore 
exotic quantum phases in the future.

\begin{acknowledgments}
G. S. is appreciative of support from the NSFC under the Grant Nos. 11704186 and 11874220.
W.-L.Y. is supported by the NSFC under the Grant No. 12174194, the startup fund (Grant No.1008-YAH20006) of
Nanjing University of Aeronautics and Astronautics, Top-notch Academic Programs Project of Jiangsu Higher Education Institutions,
and stable supports for basic institute research (Grant No. 190101).
Numerical simulations were performed on the clusters at Nanjing University of Aeronautics and Astronautics and National Supercomputing Center in Shenzhen.
\end{acknowledgments}

\bibliographystyle{apsrev4-1}
\bibliography{ref}

\end{document}